\documentclass[twocolumn,showpacs,amsmath,amssymb,prl]{revtex4}  

\usepackage{graphicx}
\usepackage{dcolumn}
\usepackage{bm}

\begin{document}

\title{Observation of Vortex Pinning in Bose-Einstein Condensates}
\author{S. Tung, V. Schweikhard, and E.~A. Cornell}
\affiliation{ JILA, National Institute of Standards and Technology,
and Department of Physics, University of Colorado, Boulder, Colorado
80309-0440}
\date{\today}

\begin{abstract}
We report the observation of vortex pinning in rotating gaseous
Bose-Einstein condensates (BEC). Vortices are pinned to columnar
pinning sites created by a co-rotating optical lattice superimposed
on the rotating BEC. We study the effects of two types of optical
lattice, triangular and square. In both geometries we see an
orientation locking between the vortex and the optical lattices. At
sufficient intensity the square optical lattice induces a structural
cross-over in the vortex lattice.
\end{abstract}


\maketitle
 Some of the most appealing results from recent work in
superfluid gases have had to do with lattices, either optical
lattices \cite{Greiner,Weiss,Esslinger} or vortex lattices
\cite{Darlibard,JILA,Ketterle}. These two kinds of lattices could
hardly be more different.   The former is an externally imposed
periodic potential arising from the interference of laser beams,
while the latter is the self-organized natural response of a
superfluid to rotation. As distinct as these two periodic structures
may be, there are reasons for trying to marry them in the same
experiment.  For one thing, the extreme limits of rapid rotation (in
the case of vortex lattices) \cite{Gunn} and deep potentials (in the
case of optical lattices) \cite{Jaksch} both lead to the same thing:
correlated many-body states.  For another, there is considerable
precedent, from various subdisciplines of physics, for interesting
effects arising from the interplay between competing lattices
\cite{Stephens,Bak,Goto}. Moreover, the pinning of superconducting
flux vortices to an array of pinning sites in solids is an area of
very active research as well \cite{scvortex}. With these
considerations in mind, we undertook a preliminary experimental
study of the effects of a rotating optical lattice on a vortex
lattice in a Bose-condensed sample of $^{87}$Rb. The density of the
superfluid is suppressed at the antinodes of the two-dimensional
standing wave pattern of the optical lattice. These antinodes then
become pinning sites, regions of low potential energy, for the
superfluid vortices. Vortices can lower their interaction energy by
arranging themselves to be as far apart as possible from one
another. The competition between these effects has been examined in
several theoretical works \cite{Reijnders,Pu}. Also \cite{Wu,Rajiv}
discuss similar systems in the strong interacting regime
\par
\begin{figure}
  \includegraphics[width=3in]{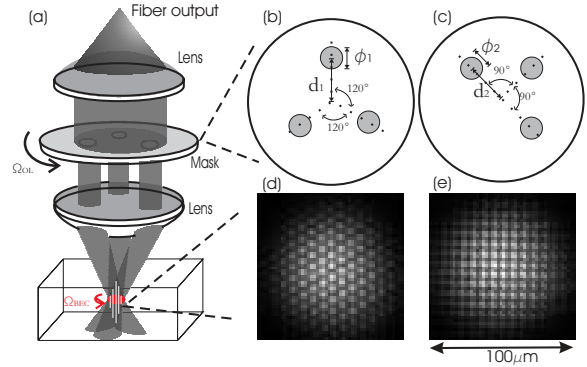}
  \caption{(a) Schematic diagram of our setup for the rotating quasi-2D optical lattice.
  Layouts of the masks for a triangular (b) and square (c) optical lattices.
  (d) and (e) are pictures of triangular and square optical lattices, respectively.  For
  details of the layouts see \cite{mask}.}\label{setup}
\end{figure}

The setup for creating a rotating optical lattice is shown in
Fig.~\ref{setup}(a). A mask with a set of holes is mounted onto a
motor-driven rotary stage, and a laser beam (532 nm) is expanded,
collimated, and passed through the mask. After the mask the
resulting three beams are focused
 onto the Bose-Einstein condensates (BEC). The interference pattern at the focus constructs a
quasi-2D optical lattice. The geometry and spatial extent of the
triangular or the square optical lattice is determined by the size
and layout of the holes and the focal length of the second lens.
For the pinning sites to appear static in the frame of a rotating
BEC, the rotation of the two lattices must be concentric, and
mechanical instabilities and optical aberrations (which lead to
epicyclic motion of the pinning sites) must be particularly
minimized. Even so, residual undesired motion is such that the
strength of the optical lattice must be kept at less than 30$\%$ of
the condensate's chemical potential or unacceptable heating results
over the experiment duration of tens of seconds. We work perforce in
the weak pinning regime.
\par
The experiments begin with condensates containing $\sim 3 \times
10^6$ $^{87}$Rb atoms, held in the Zeeman state
$|F=1,m_{f}=-1\rangle$ by an axial symmetric magnetic trap with
trapping frequencies $\{\omega_r,\omega_z\}=2\pi\{8.5,5.5\}$Hz.
Before the optical lattice, rotating at angular frequency
$\Omega_{OL}$, is ramped on, the BEC is spun up \cite{JILA} close to
$\Omega_{OL}$. This leads, before application of an optical lattice,
to the formation of a near perfect triangular vortex lattice with a
random initial angular orientation in inertial space. Through
dissipation a vortex lattice can come to equilibrium with an optical
lattice, with their rotation rates and angular orientations locked.
In the absence of pinning sites, a vortex lattice with areal density
of vortices $n_v$ will rotate at (approximately
\cite{Ian,Radzihovsky}) $\Omega=(\frac{\hbar \pi}{m})n_{v}$. This
suggests that for an optical lattice with an areal density of
pinning sites $n_{OL}$, locking between the two lattices will be
facilitated if the optical lattice rotates at the commensurate
frequency $\Omega_c \equiv (\frac{\hbar \pi}{m})n_{OL}$.

\par
\begin{figure}
  \includegraphics[width=3.0in]{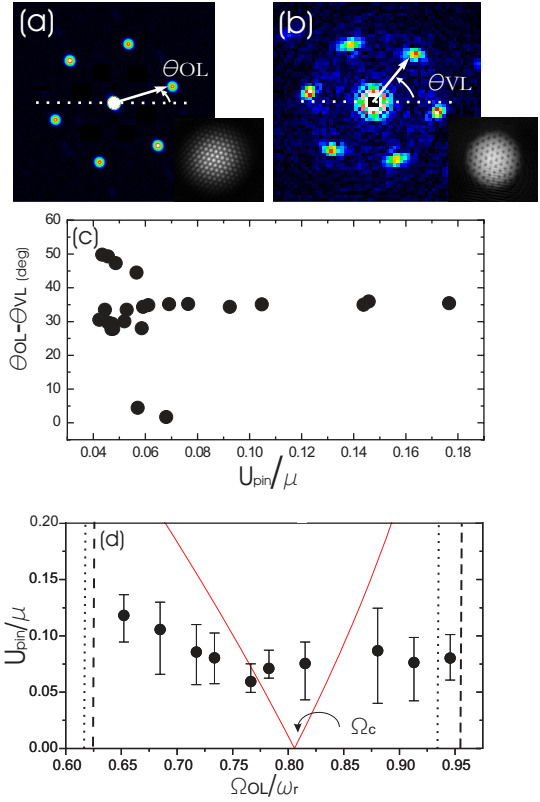}
  \caption{(a) Triangular optical lattice and (b) vortex lattice in reciprocal space. Each inset shows the corresponding original real-space
  CCD-camera images. (c) The difference in orientation $\theta_{OL}-\theta_{VL}$ versus the strength of pinning $U_{pin}/\mu$
  ( the peak of the optical potential normalized by the condensate's chemical potential) for the rotation rates
  $\Omega_{OL}= 1.133 \Omega_{c}= 0.913 \omega_{r}$. With increasing pinning strength, $\theta_{OL}-\theta_{VL}$ tends towards its locked value \cite{offset}.
  (d) Minimum pinning strength needed for orientation locking between two lattices as a function of the rotation rate of the optical lattice. The
  dashed and dotted lines are discussed in the text.}\label{anglock}
\end{figure}

We measure the angular difference $\theta_{OL}-\theta_{VL}$ between
the orientation of the optical and vortex lattice in reciprocal
space (see Figs.~2(a)--2(b)). Fig.~\ref{anglock}(c) shows
$\theta_{OL}-\theta_{VL}$ as a function of the pinning strength with
an optical lattice rotation rate $\Omega_{OL}= 1.133 \Omega_{c} =
0.913 \omega_{r}$. The strength of pinning is characterized by the
ratio $U_{pin}/\mu$ ($\mu$ is the chemical potential of the
condensate \cite{chemical}), which gives the relative suppression of
the superfluid density at pinning sites. We can see the initially
random angular difference between the two lattices becomes smaller
as the pinning strength $U_{pin}/\mu$ increases. For $U_{pin}/\mu
\gtrsim 0.08$ , the angular differences become very close to the
locked value.
Figure~\ref{anglock}(d) shows the phase diagram. The data points and
error bars mark the minimum pinning strength $(U_{pin}/\mu)_{min}$
above which the lattices lock. We observe two distinct regimes.
First, for small rotation-rate mismatch, $(U_{pin}/\mu)_{min}$ is
rather independent of the rotation-rate mismatch. Second, for
rotation-rate mismatch beyond the range indicated by the dashed line
in Fig.~\ref{anglock}(d), angular orientation locking becomes very
difficult for any $U_{pin}/\mu$ in our experiment. Instead, an
ordered vortex lattice with random overall angular orientation
observed at low $U_{pin}/\mu$ transforms into a disordered vortex
arrangement at high $U_{pin}/\mu$.
\par
This box--like shape of the locked region in $U_{pin}-\Omega_{OL}$
space is worth considering. In a simple model, vortex motion in our
system is governed by a balance of the pinning force and the Magnus
force. The pinning force is $\overrightarrow{F}_{pin}(x)\propto
U_{pin}/d$, where $U_{pin}$ and $d$ are the strength of the pinning
potential and its period, respectively. The Magnus force, acting on
a vortex moving with velocity $\overrightarrow{v}_{vortex}$ in a
superfluid with velocity $\overrightarrow{v}_{fluid}$ is
$\overrightarrow{F}_{mag}(x)\propto
n(x)\,(\overrightarrow{v}_{vortex}-\overrightarrow{v}_{fluid})\times
\overrightarrow{\kappa}$ where $\overrightarrow{\kappa}=
(\frac{h}{m})\hat{z}$, and $n(x)$ is the superfluid density. A
locked vortex lattice will co-rotate with the pinning potential,
giving
$\overrightarrow{v}_{vortex}(r)=\overrightarrow{v}_{OL}(r)=\overrightarrow
{\Omega}_{OL}\times \overrightarrow{r}$, whereas the superfluid
velocity in a solid-body approximation is
$\overrightarrow{v}_{fluid}(r)=\frac{\hbar
\pi}{m}n_{v}r\hat{\theta}=\overrightarrow{\Omega}_{fluid}\times
\overrightarrow{r}$. Comparing the magnitudes of both forces at
$r=R(\Omega)/2$, where $R(\Omega)$ is the centrifugal-force modified
Thomas-Fermi radius, we obtain a minimum strength for pinning
$(U_{pin}/\mu)_{min} \approx (\frac{1}{2\sqrt{3}} R{(\Omega)}/d)
\times(\Omega_{OL}-\Omega_{fluid})/\Omega_{c}$.
\par
What will be the fluid rotation rate $\Omega_{fluid}$ in the
presence of the pinning potential? On the one hand, if vortices are
tightly locked to the optical lattice sites, we have $\Omega_{fluid}
= \Omega_{c}$. The minimum strength $(U_{pin}/\mu)_{min}$ resulting
from this assumption is plotted as solid line in
Fig.~\ref{anglock}(d). The lack of predicted decrease of
$(U_{pin}/\mu)_{min}$ to zero around $\Omega_{c}$ may be due to long
equilibration times in a very shallow pinning potential, as well as
slight mismatches in alignment and initial rotation rate of the BEC
and the pinning potential. The ease of orientation locking with
increasing rotation rate mismatch is less easy to explain in this
model. On the other hand, in the weak-pinning regime, the vortex
lattice can accommodate a rotation rate mismatch by
stretching/compressing away from the pinning sites. This allows the
fluid to co-rotate with the optical lattice ($\Omega_{fluid} \approx
\Omega_{OL}$) and reduce the Magnus force. This leads to a very low
minimum pinning strength, as suggested by our data. However, the
vortex lattice's gain in pinning energy decreases rapidly in the
locked orientation when the mismatch between vortex spacing and
optical lattice constant increases to the point where the outermost
vortices fall radially in between two pinning sites. Then the
preference for the locked angular orientation vanishes. This
predicted limit is indicated by the vertical dotted lines in
Fig.~\ref{anglock}(d).

\par
\begin{figure}
  \includegraphics[width=3in]{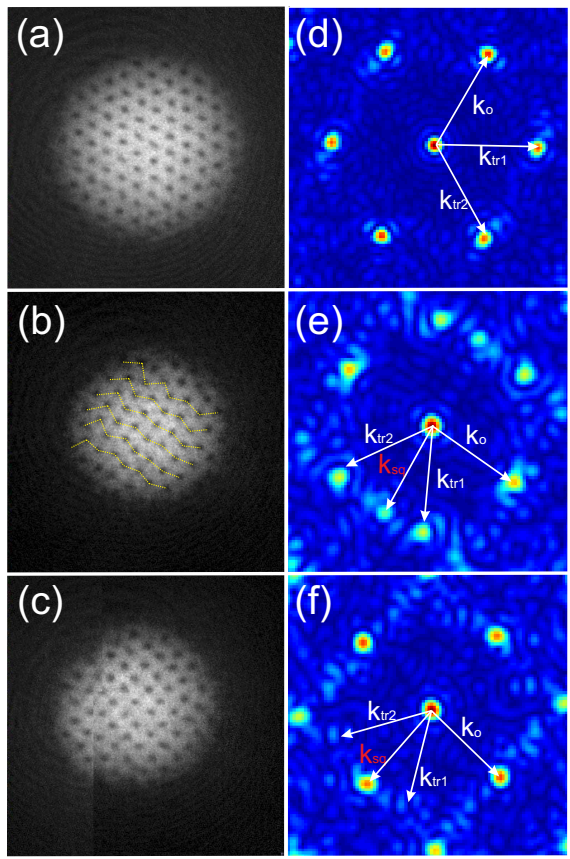}
  \caption{Images of rotating condensates pinned to an optical lattice at $\Omega_{OL}=\Omega_{c}= 0.866 \omega_{r}$
  with pinning strength $U_{pin}/\mu$=(a) 0.049 (b) 0.084 (c) 0.143, showing the structural cross-over of the
  vortex lattice. (a)--(c) are the absorption images of the vortex lattices after
  expansion. (d)--(f) are the Fourier transforms of the images in (a)--(c). $k_{o}$ is taken by convention to be the strongest peak;
  $k_{tr1}$, $k_{sq}$, and $k_{tr2}$ are at $60^{\circ}$, $90^{\circ}$, and $120^{\circ}$, respectively, from $k_{o}$.}\label{SPT}
\end{figure}
\begin{figure}
  \includegraphics[width=2.7in]{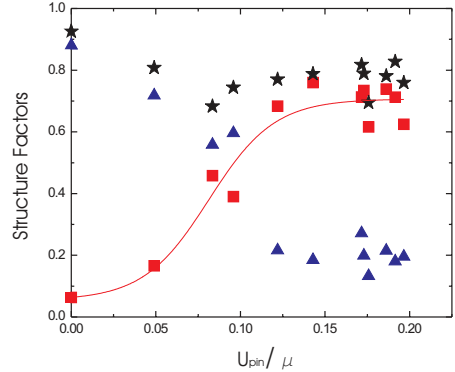}
  \caption{Structure factors (a) $|S(k_{sq})|$ ($\blacksquare$), (b) $|S(k_{tr})|$ ($\blacktriangle$)
  (average of $|S(k_{tr1})|$ and $|S(k_{tr2})|$), and (c) $|S(k_{o})|$ $(\bigstar)$ versus the strength of the
  pinning lattice at the commensurate rotation rate $\Omega_c$. $|S(k_{sq})|$ is fitted by \cite{Boltzman}.
  The fitting leads to a maximum value 0.707 of $|S(k_{sq})|$. An ideal square vortex lattice would have $|S(k_{sq})|$=1.}\label{sfactor}
\end{figure}
In the absence of a pinning potential, the interaction energy of a
square vortex lattice is calculated to exceed that of a triangular
lattice by less than 1$\%$ \cite{Campbell}, thus it is predicted
\cite{Reijnders,Pu} that the influence of even a relatively weak
square optical lattice will be sufficient to induce a structural
transition in the vortex lattice. This structural cross-over of a
vortex lattice is observed in our experiment. Figure~\ref{SPT} shows
how the vortex lattice evolves from triangular to square as the
pinning strength increases. Over a wide range of pinning strengths,
we observe that there is always at least one lattice peak in
reciprocal space that remains very strong. We define this peak to be
$k_{o}$. Lattice peaks at $60^{\circ}$ and $120^{\circ}$ from
$k_{o}$ are referred to as $k_{tr}$, and, together with $k_{o}$,
their strength is a measure of the continued presence of a
triangular lattice.  A peak at $90^{\circ}$, referred to as
$k_{sq}$, is instead a signal for the squareness of the vortex
lattice. With increasing pinning strength (Fig.~\ref{SPT}(a--c), or
(d--f)), we see the triangle to square crossover evolve. At
intermediate strengths ($U_{pin}/\mu$ = 0.084), a family of zigzag
vortex rows emerges, indicated by the dotted lines in
Fig.~\ref{SPT}(b); in reciprocal space we see the presence of
structure at $k_{tr}$ and $k_{sq}$.
\par
\begin{figure}
  \includegraphics[width=2.7in]{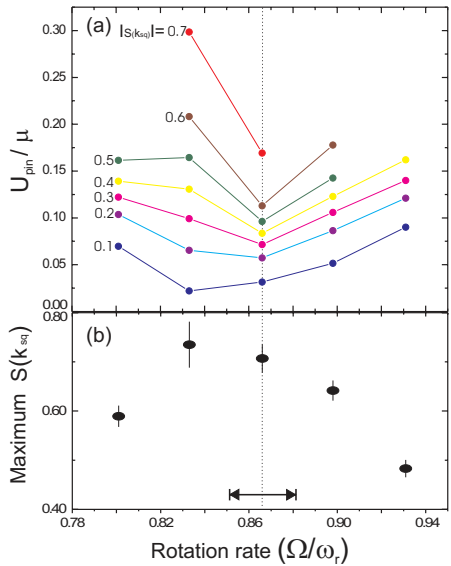}
  \caption{Effect of square pinning lattice. (a) Contours of
  $|S(k_{sq})|$ are plotted, showing the effect of the rotation
  rate and pinning strength on the squareness of the vortex lattice. (b) The maximum observed squareness. In (a) and (b), for each rotation rate, the
  data points are extracted from fits such as that shown in Fig.~\ref{sfactor}
  for $\Omega_{OL} = \Omega_{c}$. The vertical dotted line plus arrow shows the possible range
  of $\Omega_{c}$ consistent with the uncertainty in $n_{OL}$.}\label{sqphase}
\end{figure}

We quantify the crossover by means of an image-processing routine
that locates each vortex core, replaces it with a point with unit
strength, Fourier transforms the resultant pattern, and calculates
structure factors $|S|$ \cite{Pu} based on the strength of the
images at lattice vectors $k_{sq}$, $k_{tr}$, and $k_{o}$.  In
Fig.~\ref{sfactor}, we see with increasing optical potential the
turn-on of $|S(k_{sq})|$ balanced by the turn-off of $|S(k_{tr})|$.
We use a fitting function to smooth the noisy data of $|S(k_{sq})|$.
The structure crossover takes place around $U_{pin}/\mu\approx$
8$\%$, in rough agreement with predictions of $U_{pin}/\mu\approx$
5$\%$ from numerical simulations \cite{Pu} and $U_{pin}/\mu\approx$
1$\%$ from analytic theory for infinite lattices \cite{Reijnders}.
The fact that one lattice peak remains strong for all pinning
strengths (the stars ($\bigstar$) in Fig. 4) suggests that as the
pinning strength is increased, one family of vortex rows represented
by $k_{o}$ in Fourier space locks to the square pinning lattice and
remains locked as the shape cross-over distorts the other two
families of vortex rows into a square geometry. The effects of
various rotation rates and optical potential strengths on the
squareness of the vortex lattice is summarized in
Fig.~\ref{sqphase}. We surmise that there are a number of effects at
play. When $\Omega_{OL}$ differs from $\Omega_{c}$, pinning strength
is required not only to deform the shape of the vortex lattice from
triangular to square, but also to compress or expand it to match the
density of the optical lattice sites. At higher optical intensities,
we know from separate observations that imperfections in the
rotation of the optical lattice lead to heating of the condensate,
which may limit the obtainable strength of the square lattice.

\par
\begin{figure}
  \includegraphics[width=2.3in]{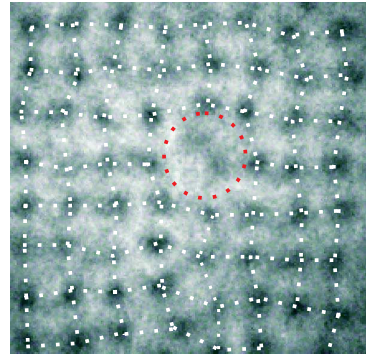}
  \caption{Image of a dumbbell-shape defect consisting of two vortices
  locked to one pinning site during the formation
  of the square vortex lattice. Dotted lines are to guide the eye.}
  \label{defects}
\end{figure}

A dumbbell-shape lattice defect (Fig.~\ref{defects}) is sometimes
observed in the early stages of the square vortex lattice formation
when $\Omega_{OL}>\Omega_{c}$. In the weak-pinning regime, the
defect will relax towards the equilibrium configuration by pushing
extra vortices at the edge of the condensate outside the system.
Defects of this nature, involving extra (or missing) vortices, are
the exception and not the rule in our observations, even for
$\Omega_{OL}\neq\Omega_{c}$. In an infinite system, the physics of
the lattice-lattice interaction would likely be dominated by these
point defects. In our finite system, would-be incommensurate
lattices can accommodate by stretching or compressing.

\par
The work presented in this paper was funded by NSF and NIST. We
would like to acknowledge P. Engels and M. Friedman for help with
the early experimental setup. We would also like to thank A.
Leanhardt, D. Sheehy, M. Kraemer, and R. Bhat for useful discussion
and H. Green for helping manufacture the rotating masks.




\begin{thebibliography}{20}
\bibitem{mask}
Mask (b): $\phi_1$= 2.5 mm and $d_1$= 11.5 mm. The lattice constant
of the triangular optical lattice is 7.8 $\mu$m, giving
$\Omega_{c}$= 0.806 $\omega_{r}$.; Mask (c): $\phi_2= 2.5$ mm and
$d_2$= 13.5 mm. The lattice constant of the square lattice is 7.0
$\mu$m, giving $\Omega_{c}$= 0.866 $\omega_{r}$. The focal lens of
two lenses are 300 mm and 250 mm, respectively.
\bibitem{offset}
There is a constant measurement offset to the values of
$\theta_{OL}-\theta_{VL}$ in this data set. We have used in-trap
images of vortex lattices to confirm that the absolute value of
$\theta_{OL}-\theta_{VL}$ is zero for locked lattices
\bibitem{chemical}
The chemical potential is determined from the axial Thomas-Fermi
radius of the condensate obtained from in-trap images, and is
proportional to the peak density of the condensate.
\bibitem{Boltzman}
$|S(k_{sq})|$=$A_2+\{(A_1-A_2)/(1+Exp[(U-U_0)/(\delta U)])\}$;
$A_1$, $A_2$, $U_0,$ and $\delta U$  are fit to the data.
\end{thebibliography}
\end{document}